\documentclass[letterpaper,twocolumn,showpacs,prl,aps]{revtex4-1}

\usepackage{amsmath}  
\usepackage{amsfonts} 
\usepackage{graphicx} 
\usepackage{amssymb}

\begin{document}

\title{Kelvin-Mach wake in a two-dimensional Fermi sea}

\author{Eugene B. Kolomeisky$^{1}$ and Joseph P. Straley$^{2}$}

\affiliation
{$^{1}$Department of Physics, University of Virginia, P. O. Box 400714,
Charlottesville, Virginia 22904-4714, USA\\
$^{2}$Department of Physics and Astronomy, University of Kentucky,
Lexington, Kentucky 40506-0055, USA}

\date{\today}

\begin{abstract}
The dispersion law for plasma oscillations in a two-dimensional electron gas in the hydrodynamic approximation interpolates between $\Omega \propto \sqrt{q}$ and $\Omega \propto q$ dependences as the wave vector $q$ increases.   As a result, downstream of a charged impurity in the presence of a uniform supersonic electric current flow, a wake pattern of induced charge density and potential is formed whose geometry is controlled by the Mach number $M$.  For $1<M\leqslant \sqrt{2}$ the wake consists of transverse wavefronts confined within a sector whose angle is given by the classic Mach condition.  An additional wake of larger angle resembling the Kelvin ship wake and consisting of both transverse and diverging wavefronts is found outside the Mach sector for $M>\sqrt{2}$.  These wakes also trail an external charge traveling supersonically a fixed distance away from the electron gas.    
\end{abstract}


\maketitle

An object uniformly moving relative to a medium gives rise to a series of effects ranging from formation of a Mach shockwave cone behind a supersonic projectile \cite{LL6} and Cherenkov radiation emitted by a rapidly moving charge \cite{LL8}, to creation of wakes on water surface by ships \cite{Lamb}.  One feature these effects have in common is that the interaction between the object and the medium triggers the coherent emission of collective excitations of the medium which combine constructively to form the wake \cite{wake_review}.  Here we describe the coherence effect wherein plasma waves emitted by a two-dimensional ($2d$) electron gas form a wake pattern resembling both Mach and ship wakes \cite{Lamb}.  Hereafter we speak of the electron gas;  the theory for the gas of holes is the same.        

The starting point of our analysis is an expression for the dynamical dielectric function of the $2d$ electron gas in the hydrodynamic approximation which, neglecting the effects of dissipation and retardation, is given by \cite{AFS,Fetter}: 
\begin{equation}
\label{dyn_diel_function}
\epsilon(\omega,\textbf{q})=\frac{\omega^{2}-\Omega^{2}(\textbf{q})}{\omega^{2}-s^{2}q^{2}}
\end{equation}
where $\omega$ is the frequency, $\Omega(\textbf{q})$ is the frequency of plasma oscillations as a function of the wave vector $\textbf{q}$,  
\begin{equation}
\label{spectrum}
\Omega^{2}(\textbf{q})=gq+s^{2}q^{2}
\end{equation}
and $q=|\textbf{q}|$.  This description encompasses systems ranging from those whose electrons obey parabolic \cite{AFS, Fetter} to linear (graphene) dispersion laws \cite{DasHwang}. The  material parameters $g=2\pi n e^{2}v_{F}^{2}/\kappa \zeta(n)$ (a characteristic acceleration) and $s=v_{F}(\partial p/\partial \varepsilon)^{1/2}$ (the speed of sound) are determined by the equilibrium electron number density $n$, the equation of state in the neutral limit (entering via the density dependence of the chemical potential $\zeta(n)$ and the energy density dependence of the pressure $p(\varepsilon)$), the background dielectric constant $\kappa$, and the limiting (Fermi) velocity $v_{F}$ \cite{KS}.

In the long-wavelength limit $\textbf{q}\rightarrow 0$ the spectrum (\ref{spectrum}) is formally the same as that of gravity waves on deep water, $\Omega^{2}(\textbf{q})=gq$ \cite{Lamb} (where in this context $g$ is the free-fall acceleration), an observation due to Dyakonov and Shur \cite{DS}.  Since plasma oscillations are classical in nature \cite{KS}, a series of effects analogous to classical waves on water are then expected in electron layers.  

One of the most familiar manifestations of the $\Omega(\textbf{q}\rightarrow 0)\propto \sqrt{q}$ dispersion law in fluid mechanics is the Kelvin wake that trails a traveling pressure source:  the $39^{\circ}$ angle of the wake is independent of the source velocity and has a characteristic "feathered" pattern \cite{Lamb}.  One then might infer that an external charge traveling non-relativistically a fixed distance away from the plane of the electron system disturbs the latter in the form of an "electron"  Kelvin wake.   Such a conclusion was recently made in the literature in the context of doped graphene \cite{proposal};  it is misleading because it overlooks crucial deviation from the strict $\sqrt{q}$ dispersion law.  The same criticism applies to a conjecture that stationary Kelvin wake should be formed downstream of a defect in the $2d$ electron gas in the presence of a current \cite{Dyakonov16}.  A wake is formed behind a moving source whenever there is a mode whose phase velocity matches the speed of the source (the precise statement is given by Eq.(\ref{Cherenkov_Landau}) below).  For a strictly $\sqrt{q}$ dispersion law such a mode can always be found no matter what the speed of the source.  However, the spectrum of plasma oscillations (\ref{spectrum}) deviates from the $\sqrt{q}$ law, and the phase velocity $\Omega/q$ is always above the speed of the sound $s,$ which is thus the critical velocity for wake formation in $2d$ electron systems.  If the acceleration $g$ in Eq.(\ref{spectrum}) were zero, the wake pattern would resemble that formed behind a supersonic projectile, with a wake angle determined by the Mach number $M=v/s$ \cite{LL6}, where $v$ is the speed of the source.  For finite Mach number one would then expect a pattern sharing features of both the Kelvin and Mach wakes, hereafter called the Kelvin-Mach wake.  

In an earlier study, Fetter \cite{Fetter} has analyzed various aspects of the electromagnetic response of an electron layer to a moving charge.  However, the problem was solved in the Fourier representation, and the real space pattern of the induced charge and potential were not addressed.  

Our goal is to solve for the geometry of the wake induced by the moving charge.  This is done by focusing on the case when the external charge is in the plane of the electron system.  Since only the relative motion of the charge and the medium matters, in practice this situation can be realized by subjecting an electron layer with an embedded Coulomb impurity to a supersonic current flow.  

Supersonic flows are experimentally accessible, as we will now show.  

*The speed of sound $s$ is less than the Fermi velocity $v_{F}$ but typically has the same order of magnitude.

*For a parabolic dispersion law it can be estimated as $s\simeq v_{F}\simeq c (m_{0}/137m)a_{B}\sqrt{n}$ where $m_{0}$ is the electron mass in vacuum and $a_{B}$ is the Bohr radius.  For $m_{0}/m=10$ and $n=10^{12}cm^{-2}$ the speed of sound can be estimated as $s\simeq 10^{6} cm/s$.   This large value can be attained at low temperature, where the mobility can be as large as $10^{4} cm^{2}/(V\cdot s)$ \cite{AFS}.  The required electric field would be $10^{2}V/cm$, which is five orders of magnitude smaller than the dielectric breakdown field of the $SiO_{2}$ insulating layer common to various practical realizations of electron layer systems \cite{AFS}.  

*Similarly, in graphene (a linear dispersion material) the Fermi velocity is two orders of magnitude smaller than the speed of light but the electron mobility is of the order $10^{5} cm^{2}/(V\cdot s)$ \cite{graphene_review} which translates into a $10^{3}V/cm$ electric field needed to propel graphene's electrons past the speed of sound.  There exists direct experimental evidence \cite{direct} that the saturation velocity in graphene on $SiO_{2}$ above room temperature exceeds $3\times 10^{7}cm/s$ at low carrier density while the intrinsic graphene saturation velocity could be more than twice the quoted value.  

*The ratio $d =s^2/g$ (the Debye screening length of the electron gas \cite{AFS}) sets the length scale of the effects to be discussed.   It is of the order $10^{-6} cm$ (and weakly doping dependent) in materials with parabolic dispersion law \cite{AFS} and of the order $\kappa/\sqrt{n}$ in graphene \cite{graphene_review}.

*There is a further advantage of studying graphene rather than the electron layers of the past \cite{AFS}.  Charged impurities can be embedded into graphene in a controlled manner \cite{Eva_Andrei} and high-resolution non-invasive imaging of charge currents in graphene structures \cite{imaging} can be employed to directly observe the electron Kelvin-Mach wake;  in other systems the formation of the wake can only be inferred indirectly from the onset of non-zero wave resistance.       

We will be studying the electromagnetic response of an electron layer to an external potential $\varphi_{ext}(\textbf{r},t)$, where $\textbf{r}$ is the position within the layer and $t$ is the time; the dependence on these quantities is in respose to an  external charge (number) density $n_{ext}(\textbf{r},t)$.  Their Fourier transforms are related by the Coulomb law $\varphi_{ext}(\omega,\textbf{q})=2\pi en_{ext}(\omega,\textbf{q})/\kappa q$ \cite{AFS,Fetter}.  According to the linear response theory, the Fourier components of the induced density $n_{in}(\omega,\textbf{q})$ and induced potential $\varphi_{in}(\omega,\textbf{q})$ are given by
\begin{equation}
\label{linear_response_Fourier_density}
n_{in}(\omega,\textbf{q})=\left [\frac{1}{\epsilon(\omega,\textbf{q})}-1\right ]n_{ext}(\omega,\textbf{q})
=\frac{gqn_{ext}(\omega,\textbf{q})}{\omega^{2}-\Omega^{2}(\textbf{q})}
\end{equation} 
\begin{equation}
\label{linear_response_Fourier_potential}
\varphi_{in}(\omega,\textbf{q})=\left [\frac{1}{\epsilon(\omega,\textbf{q})}-1\right ]\varphi_{ext}(\omega,\textbf{q})
=\frac{2\pi eg}{\kappa}\frac{n_{ext}(\omega,\textbf{q})}{\omega^{2}-\Omega^{2}(\textbf{q})}
\end{equation} 
Inverting the Fourier transforms we find the electromagnetic response in the direct space and time representation 
\begin{equation}
\label{linear_response_real_space_charge}
n_{in}(\textbf{r},t)=g\int\frac{d^{2}qd\omega}{(2\pi)^{3}}\frac{qn_{ext}(\omega,\textbf{q})e^{i(\textbf{q}\cdot\textbf{r}-\omega t)}}{(\omega+i0)^{2}-\Omega^{2}(\textbf{q})}
\end{equation}
\begin{equation}
\label{linear_response_real_space_potential}
\varphi_{in}(\textbf{r},t)=\frac{2\pi eg}{\kappa}\int\frac{d^{2}qd\omega}{(2\pi)^{3}}\frac{n_{ext}(\omega,\textbf{q})e^{i(\textbf{q}\cdot\textbf{r}-\omega t)}}{(\omega+i0)^{2}-\Omega^{2}(\textbf{q})}
\end{equation}
where $\omega$ in the denominators of the integrands is endowed with infinitesimally small positive imaginary part ($\omega\rightarrow \omega+i0$) to guarantee analyticity of the integrands in the upper half-plane of complex $\omega$ \cite{LL8}.  A unit external charge moving with constant velocity $\textbf{v}$ within the layer is described by $n_{ext}(\textbf{r},t)=\delta(\textbf{r}-\textbf{v}t)$ whose Fourier transform is $n_{ext}(\omega,\textbf{q})=2\pi \delta(\omega-\textbf{q}\cdot\textbf{v})$.  Substituting this into Eqs.(\ref{linear_response_real_space_charge}) and (\ref{linear_response_real_space_potential}) and changing the frame of reference to that of the charge, $\textbf{r}-\textbf{v}t\rightarrow \textbf{r}$, we find 
\begin{equation}
\label{2d_ship_waves_general_charge}
n_{in}(\textbf{r})=g\int\frac{d^{2}q}{(2\pi)^{2}}\frac{qe^{i\textbf{q}\cdot\textbf{r}}}{(\textbf{q}\cdot\textbf{v}+i0)^{2}-\Omega^{2}(\textbf{q})}
\end{equation} 
\begin{equation}
\label{2d_ship_waves_general_potential}
\varphi_{in}(\textbf{r})=\frac{2\pi eg}{\kappa}\int\frac{d^{2}q}{(2\pi)^{2}}\frac{e^{i\textbf{q}\cdot\textbf{r}}}{(\textbf{q}\cdot\textbf{v}+i0)^{2}-\Omega^{2}(\textbf{q})}
\end{equation} 
This is the electrodynamic response of the electron layer having an initially uniform flow velocity $-\textbf{v}$ to a point Coulomb impurity of unit charge fixed at the origin or, equivalently, to a traveling charge in the co-moving reference frame. 

For $\textbf{v}=0$ Eqs.(\ref{2d_ship_waves_general_charge}) and (\ref{2d_ship_waves_general_potential}) describe the static screening response of the electron layer to a point charge \cite{AFS,Fetter}.  Slow motion ($v<s$) brings anisotropy to the response but no other qualitative changes occur because the denominators of the integrands in Eqs.(\ref{2d_ship_waves_general_charge}) and (\ref{2d_ship_waves_general_potential}) cannot vanish for $\textbf{q}$ real;  the $+i0$ frequency shift in the integrands is unimportant.  This regime, where no plasma waves are emitted, will be discussed elsewhere.  However, when $v$ exceeds the speed of sound, the denominators of the integrands in Eqs.(\ref{2d_ship_waves_general_charge}) and (\ref{2d_ship_waves_general_potential}) can vanish;  integrals (\ref{2d_ship_waves_general_charge}) and (\ref{2d_ship_waves_general_potential}) are dominated by the real wave vectors $\textbf{q}$ given by the solutions to
\begin{equation}
\label{Cherenkov_Landau}
\Omega(\textbf{q})=\pm \textbf{q}\cdot \textbf{v}
\end{equation}
The response pattern is now qualitatively different and the presence of the $+i0$ shift is required to supply a rule for bypassing the poles of the integrands in Eqs.(\ref{2d_ship_waves_general_charge}) and (\ref{2d_ship_waves_general_potential}). This regime is our focus.  For the special case of a charge moving through a medium with velocity exceeding the phase velocity of light the condition (\ref{Cherenkov_Landau}) is encountered in the theory of the Cherenkov effect \cite{LL8}.  In its general form Eq.(\ref{Cherenkov_Landau}) was given by Landau as a threshold for emission of elementary excitations by a superfluid flowing along a capillary \cite{Landau}.   

The theory is linear, the source has zero range, and the wake is stationary in the reference frame of the source.  Then dimensional analysis implies that the spatial scale of the pattern can only depend on the parameters of the spectrum $g$ and $s$ (\ref{spectrum}), and velocity of the source $v$:

(i) For $s=0$ (the Kelvin wake) the only parameter having dimensions of length that can be formed out of $g$ and $v$ is the  characteristic length scale of the wake, $\lambda=v^{2}/g$.  Measuring the length in units of $\lambda$, density in units of $1/\lambda^{2}$, and potential in units of $e/\kappa\lambda$ eliminates all the parameters from the problem.  Thus all Kelvin wakes are geometrically similar.  While this argument does not supply the value of the wake angle, it does predict that it is independent of $v$ and $g$.    

(ii)  For $s\neq0$ (the Kelvin-Mach wake) two independent length scales can be formed out of the parameters of the problem: $\lambda=v^{2}/g$ and $d=s^{2}/g$ (the Debye screening length). Their ratio, $\lambda/d=v^{2}/s^{2}=M^{2}$, is the square of the Mach number; once this is fixed, either $\lambda$ or $d$ may be used to characterize the length scale of the wake.   Measuring the length in units of $d$, density in units of $1/d^{2}$) and the potential in units of $e/\kappa d$ eliminates all the parameters from the problem except for the Mach number.  Thus all the Kelvin-Mach wakes of the same Mach number $M=v/s$ are geometrically similar. 

Even though the Fourier integrals (\ref{2d_ship_waves_general_charge}) and (\ref{2d_ship_waves_general_potential}) cannot be computed in closed form, the geometry of the wake pattern can be inferred with the help of Kelvin's method of stationary phase \cite{Lamb}.  The idea is that when the phase factor $f=\textbf{q}\cdot\textbf{r}$ in the integrands in (\ref{2d_ship_waves_general_charge}) and (\ref{2d_ship_waves_general_potential}) varies rapidly with $q$, the exponentials are highly oscillatory so that contributions from various elements $d^{2}q$ cancel each other;  this is the case of destructive interference with almost zero net result.  This cancelation, however, will not occur for the wavelengths for which $f$ is stationary with respect to $\textbf{q}$ (which is additionally restricted by the Cherenkov-Landau condition (\ref{Cherenkov_Landau}));  this is the case of constructive interference.  Since the integrands of the induced charge (\ref{2d_ship_waves_general_charge}) and potential (\ref{2d_ship_waves_general_potential}) differ by a smooth factor of $q$, the two wake patterns have the same geometry.  

Let us choose the positive $x$ direction along the velocity vector $\textbf{v}$ and measure length in units of the Debye screening length $d=s^{2}/g$.  The trace of the external charge divides the plane into two regions related to one another by reflection;  without the loss of generality we can focus on the $y>0$ half-space.  Here the wake is formed by a superposition of the waves whose wave vectors have positive components, $q_{x,y}>0$.  Then the phase $f=\textbf{q}\cdot\textbf{r}$ is given by 
\begin{eqnarray}
\label{phase_general}
f&=&\frac{\Big\{2(M^{2}-1) q_{y}^{2}+1+\left [1+4(M^{2}-1) M^{2}q_{y}^{2}\right ]^{1/2}\Big\}^{1/2}}{(M^{2}-1)\sqrt{2}}x\nonumber\\
&+&q_{y}y
\end{eqnarray}              
where instead of $q_{x}$ we substituted the positive solution of the Cherenkov-Landau equation (\ref{Cherenkov_Landau}) corresponding to the plasma spectrum (\ref{spectrum}).  Direct inspection of Eq.(\ref{phase_general}) shows that the condition of stationary phase $df/dq_{y}=0$ can only be satisfied for $x<0$ which is where the wake is.  In terms of a new variable $z=[1+4(M^{2}-1) M^{2}q_{y}^{2}]^{1/2}\geqslant 1$ the expression for the phase (\ref{phase_general}) can be transformed into
\begin{equation}
\label{phase(z)}
f=\frac{(z^{2}-1)^{1/2}}{2M(M^{2}-1)^{1/2}}\left[\frac{(z-1+2M^{2})^{1/2}}{(M^{2}-1)^{1/2}(z-1)^{1/2}}x+y\right]
\end{equation}
The condition of stationary phase $f'(z)=0$ now becomes
\begin{equation}
\label{stationary_phase_general}
-\frac{y}{x}=\frac{1}{(M^{2}-1)^{1/2}}\frac{(z-1)^{1/2}(z+M^{2})}{z(z-1+2M^{2})^{1/2}}
\end{equation}
Since the phase $f$ is constant along the wavefront, Eqs.(\ref{phase(z)}) and (\ref{stationary_phase_general}) can be solved relative to $x$ and $y$ to give the equation for the wavefront in parametric form:
\begin{equation}
\label{x_parametric}
x(z)=\frac{2f(M^{2}-1)}{M}\frac{z(z-1+2M^{2})^{1/2}}{(z+1)^{3/2}}
\end{equation}
\begin{equation}
\label{y_parametric}
y(z)=-\frac{2f(M^{2}-1)^{1/2}}{M}\frac{(z+M^{2})(z-1)^{1/2}}{(z+1)^{3/2}}
\end{equation}
We now see that internal consistency of the argument requires the phase to be negative, $f<0$.

As in Kelvin's case \cite{Lamb}, the range of applicability of the method of stationary phase limits our analysis to large distances $r$ from the source which in the original units of length means $r\gg d=s^{2}/g$.  

To put the consequences of Eqs.(\ref{stationary_phase_general})-(\ref{y_parametric}) into perspective we begin with the Kelvin case $s=0$ which corresponds to $M=\infty$.  In this limit Eq.(\ref{stationary_phase_general}) simplifies to
\begin{equation}
\label{stationary_phase_Kelvin}
-\frac{y}{x}=\frac{(z-1)^{1/2}}{\sqrt{2}z}
\end{equation} 
whose right-hand side vanishes at $z=1$, $z\rightarrow \infty$ and reaches a maximum value of $1/2\sqrt{2}$ in between.  Therefore the equation of stationary phase (\ref{stationary_phase_Kelvin}) has one solution for $-y/x=0$, two solutions for $0<-y/x<1/2\sqrt{2}$ coalescing at $-y/x=1/2\sqrt{2}$, and none for $-y/x>1/2\sqrt{2}$.  The angle between the wake edges is $2\arctan(1/2\sqrt{2})\approx39^{\circ}$ which is Kelvin's classic result \cite{Lamb}.  

In order to take the Kelvin $s=0$ limit in Eqs.(\ref{x_parametric}) and (\ref{y_parametric}) we temporarily restore original units of length, $(x,y)\rightarrow (x,y)/d=(g/s^{2})(x,y)$ followed by selecting $\lambda=v^{2}/g$ as a new unit of length with the result
\begin{equation}
\label{xy_parametric_Kelvin}
x(z)=\frac{2\sqrt{2}fz}{(z+1)^{3/2}},~~~~y(z)=-\frac{2f(z-1)^{1/2}}{(z+1)^{3/2}}
\end{equation}
A series of these wavefronts is shown in Figure 1 where for the purpose of illustration we chose $f=-2\pi (l+1/2), l=0,1,2$;  the $y<0$ part of the wake is obtained by reflection.
\begin{figure}
\includegraphics[width=1.0\columnwidth, keepaspectratio]{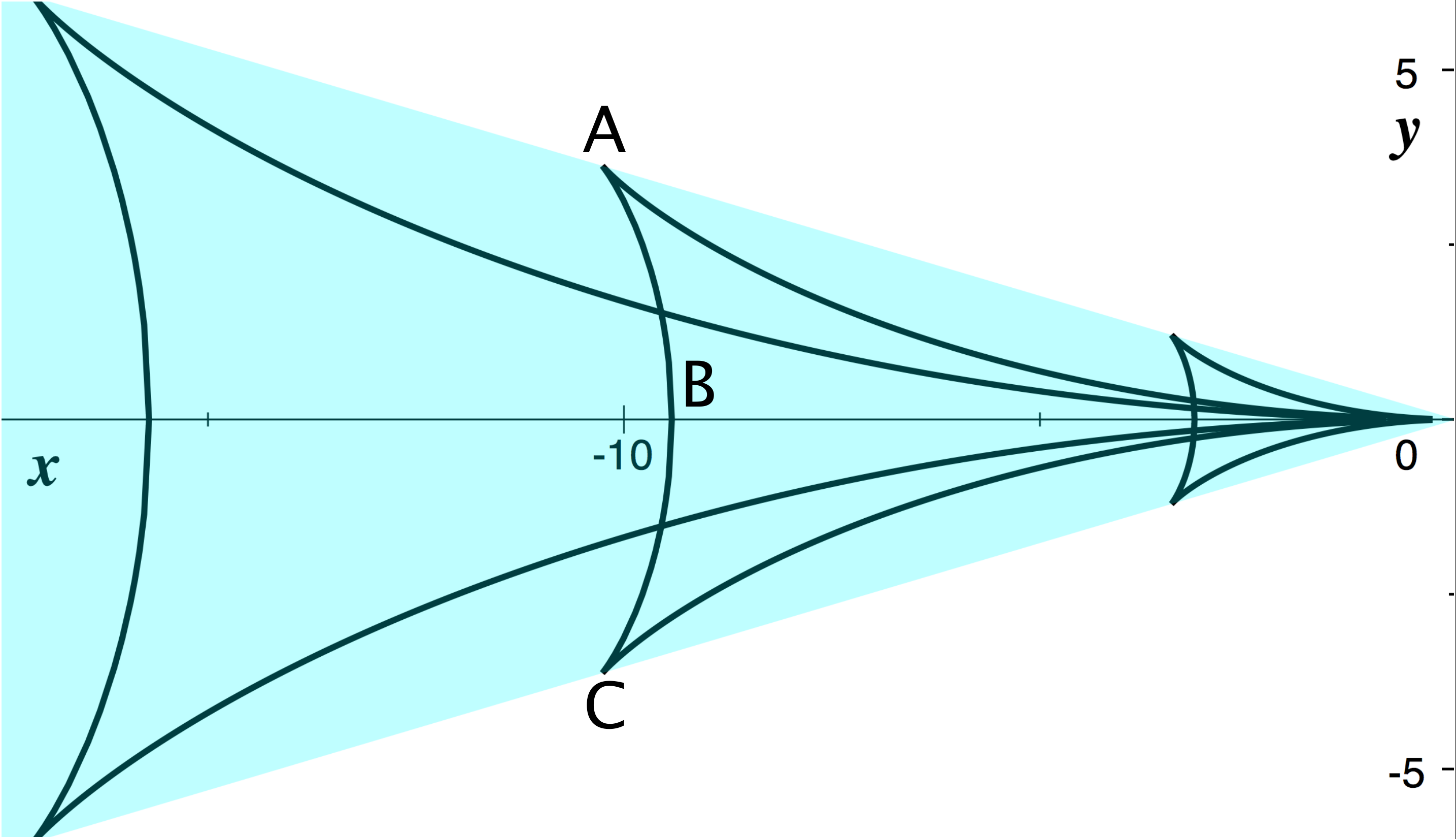} 
\caption{(Color online) Wavefronts of the Kelvin wake, Eqs. (\ref{xy_parametric_Kelvin}), with the source at the origin traveling to the right.  The wake is confined within the shaded light blue $39^{\circ}$ wedge;  the unit of length is $\lambda=v^{2}/g$.}
\end{figure} 
The wake consists of the so-called transverse wavefronts $ABC$ connecting the edges of the pattern across the central line $y=0$ and the diverging wavefronts $AO$ and $CO$ connecting the source at the origin to the edges of the pattern \cite{Lamb}.  The two wavefronts meet at $A$ and $C$ at the edges of the pattern.  
\begin{figure}
\includegraphics[width=1.0\columnwidth, keepaspectratio]{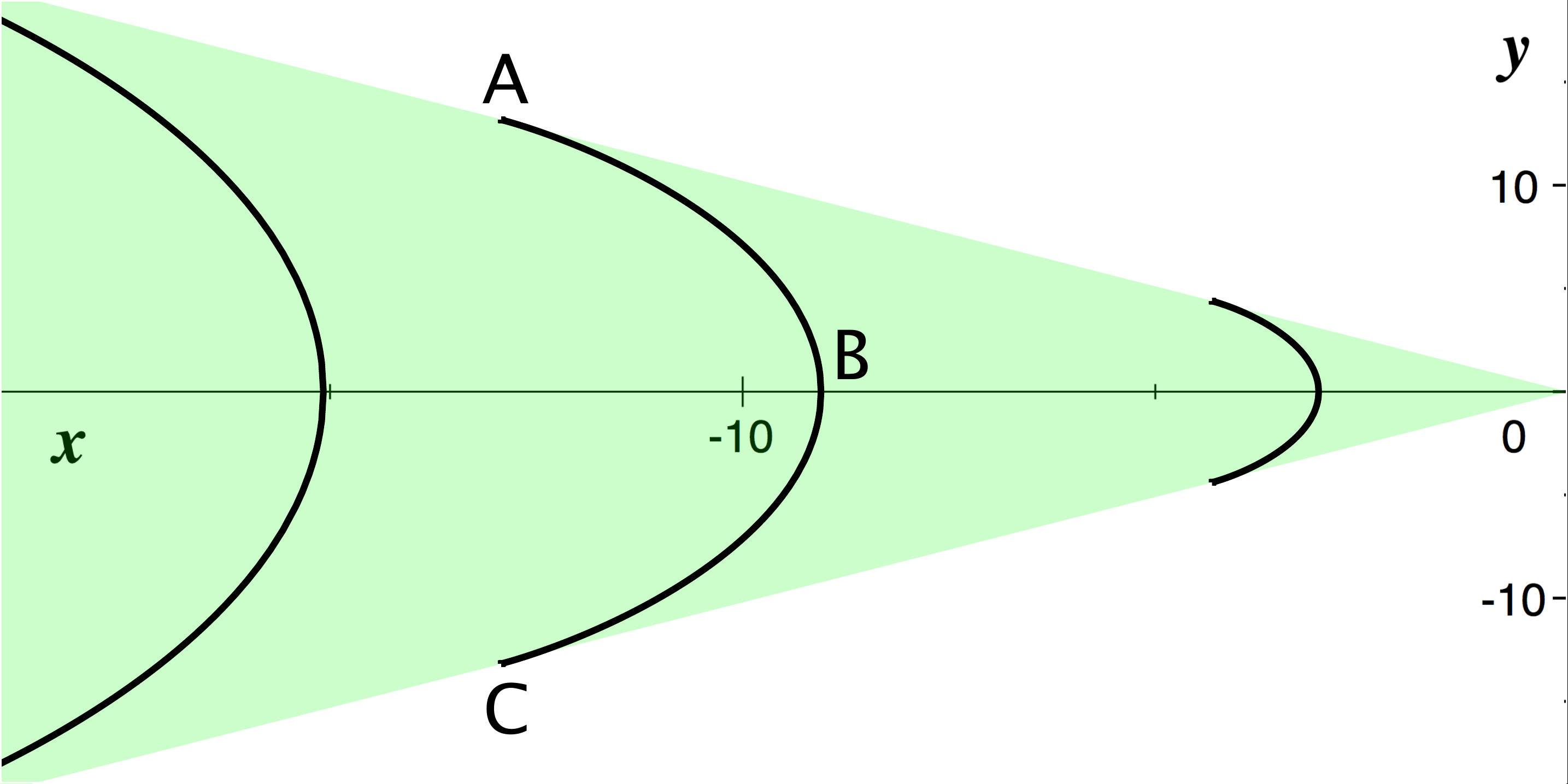} 
\caption{(Color online) Wavefronts of the Kelvin-Mach wake for $1<M\leqslant\sqrt{2}$, Eqs. (\ref{x_parametric}) and (\ref{y_parametric}), with an external charge at the origin traveling to the right.  The wake consists of transverse wavefronts confined within shaded light green Mach sector of angle $2\arctan(M^{2}-1)^{-1/2}$, the unit of length is the Debye screening length $d=s^{2}/g$, and $M=1.4$ was employed to produce the drawing.}
\end{figure}    

For $M$ finite the right-hand side of Eq.(\ref{stationary_phase_general}) vanishes at $z=1$ and approaches $(M^{2}-1)^{-1/2}$ as $z\rightarrow \infty$;  the intermediate behavior depends on the Mach number:
\begin{figure}
\includegraphics[width=1.0\columnwidth, keepaspectratio]{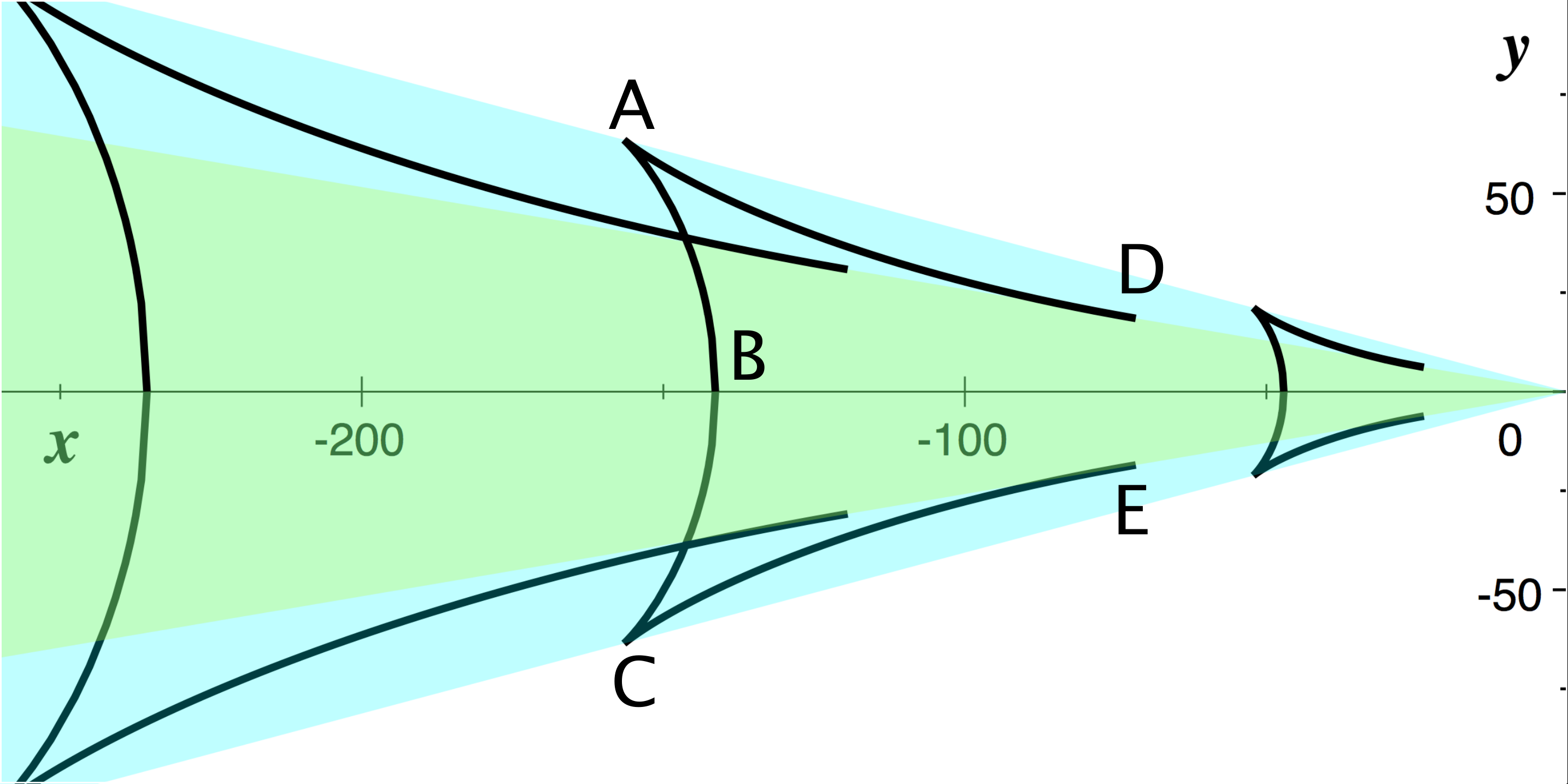} 
\caption{(Color online) Same as in Figure 2 for $M>\sqrt{2}$.  Additionally, both transverse and diverging wavefronts are present in the region shaded light blue outside the Mach sector.  The angle of the wake is given by Eq.(\ref{wider_angle }), and $M=4$ was employed to produce the drawing.}
\end{figure}

(i) When $1<M\leqslant\sqrt{2}$ the right-hand side of Eq.(\ref{stationary_phase_general}) is a monotonically increasing function of $z$.  Thus the equation of stationary phase (\ref{stationary_phase_general}) has one (transverse) solution for $0\leqslant -y/x< (M^{2}-1)^{-1/2}$  and none for $-y/x\geqslant (M^{2}-1)^{-1/2}$.  Therefore the angle of the wake is $2\arctan(M^{2}-1)^{-1/2}$ which is Mach's classic result \cite{LL6}.  A series of wavefronts (\ref{x_parametric}) and (\ref{y_parametric}) employing the same choice for the phase $f$ as in Figure 1 is shown in Figure 2.  The wake consists of transverse wavefronts $ABC$ connecting the edges of the pattern.  We stress that in view of the dispersion relation (\ref{spectrum}) the wake is \textit{not} the classic Mach wake;  the wavefronts of the latter, $y/x=\pm (M^{2}-1)^{-1/2}$, coincide with its geometrical boundary \cite{LL6}.  

(ii)  When $M>\sqrt{2}$ the right-hand side of Eq.(\ref{stationary_phase_general}) has a maximum $(M^{2}+1)^{3/2}/(2M^{2}-1)^{3/2}$ at $z=(2M^{2}-1)/(M^{2}-2)$.  Now the equation of stationary phase (\ref{stationary_phase_general}) has one (transverse) solution for $0\leqslant -y/x< (M^{2}-1)^{-1/2}$, two (transverse and diverging) solutions  for $(M^{2}-1)^{-1/2}\leqslant-y/x< (M^{2}+1)^{3/2}/(2M^{2}-1)^{3/2}$ coalescing at  $-y/x= (M^{2}+1)^{3/2}/(2M^{2}-1)^{3/2}$, and none for $-y/x> (M^{2}+1)^{3/2}/(2M^{2}-1)^{3/2}$.  The wake pattern shown in Figure 3 is confined within a sector of angle 
\begin{equation}
\label{wider_angle }
\phi(M)=2\arctan\frac{(M^{2}+1)^{3/2}}{(2M^{2}-1)^{3/2}}
\end{equation}
that is wider than Mach's.  In addition to the transverse wavefronts $ABC$ connecting the edges of the pattern, the diverging wavefronts $AD$ and $CE$ are also found outside the Mach (light green) sector;  the region with two types of wavefronts present is shaded light blue.  In contrast to the Kelvin wake (Figure 1), divergent wavefronts connect the edges of the Kelvin-Mach wake to the boundaries of the Mach sector, a consequence of the $z\rightarrow \infty$ limit of Eqs.(\ref{stationary_phase_general})-(\ref{y_parametric}).  As $M$ increases, the Mach sector becomes narrow, closing as $M\rightarrow \infty$ while the wake angle (\ref{wider_angle }) decreases approaching Kelvin's limit of $\phi(\infty)=2\arctan(1/2\sqrt{2})$.  It is expected that the appearance of the diverging wavefronts for $M>\sqrt{2}$ will be accompanied by a noticeable increase of the wave resistance. 

To summarize, our analysis has uncovered intricate wake patterns that can be produced in $2d$ electron systems which we hope will be observed in future experiments.  

We thank G. Rousseaux and M. I. Dyakonov for informing us of Refs.\cite{wake_review,Dyakonov16}, and M.I. Dyakonov and E. Y. Andrei for valuable comments.

\end{document}